\documentclass[twocolumn]{aastex631}

\usepackage[notransparent]{svg}
\usepackage{tikz}
\usepackage{amsmath}
\usepackage{xspace}

\usepackage{calligra}
\usepackage[T1]{fontenc}

\newcommand{\g}{L2\xspace}
\newcommand{\G}{L1\xspace}

\begin{document}

\title{DESI-253.2534+26.8843: A New Einstein Cross Spectroscopically Confirmed with VLT/MUSE and Modeled with GIGA-Lens}

\shortauthors{Cikota et al.}
\shorttitle{Spectroscopic confirmation and modeling of DESI-253.2534+26.8843}

\author[0000-0001-7101-9831]{Aleksandar Cikota}
\affiliation{Gemini Observatory / NSF's NOIRLab, Casilla 603, La Serena, Chile}

\author[0009-0006-3410-5531]{Ivonne Toro Bertolla}
\affiliation{Gemini Observatory / NSF's NOIRLab, Casilla 603, La Serena, Chile}

\author[0000-0001-8156-0330]{Xiaosheng Huang}
\affiliation{Department of Physics \& Astronomy, University of San Francisco, 2130 Fulton Street, San Francisco, CA 94117-1080, USA}
\affiliation{Physics Division, Lawrence Berkeley National Laboratory, 1 Cyclotron Road, Berkeley, CA 94720, USA}

\author[0009-0003-4697-7079]{Saul Baltasar}
\affiliation{Physics Division, Lawrence Berkeley National Laboratory, 1 Cyclotron Road, Berkeley, CA 94720, USA}

\author[0009-0009-8206-0325]{Nicolas Ratier-Werbin}
\affiliation{Physics Division, Lawrence Berkeley National Laboratory, 1 Cyclotron Road, Berkeley, CA 94720, USA}

\author[0000-0003-1889-0227]{William Sheu}
\affiliation{Department of Physics \& Astronomy, University of California, Los Angeles, 430 Portola Plaza, Los Angeles, CA 90095, USA}

\author[[0000-0002-0385-0014]{Christopher Storfer}
\affiliation{Institute for Astronomy, University of Hawaii, Honolulu, HI 96822-1897}

\author[0000-0001-7266-930X]{Nao Suzuki}
\affiliation{Physics Division, Lawrence Berkeley National Laboratory, 1 Cyclotron Road, Berkeley, CA 94720, USA}
\affiliation{Kavli Institute for the Physics and Mathematics of the Universe, University of Tokyo, Kashiwa 277-8583, Japan}

\author[0000-0002-5042-5088]{David J. Schlegel}
\affiliation{Physics Division, Lawrence Berkeley National Laboratory, 1 Cyclotron Road, Berkeley, CA 94720}

\author[0000-0003-4553-4033]{Regis Cartier}
\affiliation{Gemini Observatory / NSF's NOIRLab, Casilla 603, La Serena, Chile}

\author[0000-0002-2726-6971]{Simon Torres}
\affiliation{SOAR Telescope / NSF's NOIRLab, Casilla 603, La Serena, Chile}

\author[0000-0002-7671-2317]{Stefan Cikota}
\affiliation{Centro Astronónomico Hispano en Andalucía, Observatorio de Calar Alto, Sierra de los Filabres, 04550 Gérgal, Spain}

\author[0000-0002-9253-053X]{Eric Jullo}
\affiliation{Aix-Marseille Univ., CNRS, CNES, LAM, Marseille, France}

\correspondingauthor{Aleksandar Cikota, Xiaosheng Huang}
\email{aleksandar.cikota@noirlab.edu, xhuang22@usfca.edu}



\begin{abstract}
Gravitational lensing provides unique insights into astrophysics and cosmology, including the determination of galaxy mass profiles and constraining cosmological parameters. We present spectroscopic confirmation and lens modeling of the strong lensing system DESI-253.2534+26.8843, discovered in the Dark Energy Spectroscopic Instrument (DESI) Legacy Imaging Surveys data. 
This system consists of a massive elliptical galaxy surrounded by four blue images forming an Einstein Cross pattern.
We obtained spectroscopic observations of this system using the Multi Unit Spectroscopic Explorer (MUSE) on ESO's Very Large Telescope (VLT) and confirmed its lensing nature. The main lens, which is the elliptical galaxy, has a redshift of $z_{L1} = 0.636\pm 0.001$, while the spectra of the background source images are typical of a starburst galaxy and have a redshift of $z_s = 2.597 \pm 0.001$. Additionally, we identified a faint galaxy foreground of one of the lensed images, with a redshift of $z_{L2} = 0.386$.
We employed the GIGA-Lens modeling code to characterize this system and determined the Einstein radius of the main lens to be $\theta_{E} =2.520{''}_{-0.031}^{+0.032}$, which corresponds to a velocity dispersion of $\sigma$ = 379 $\pm$ 2 km s$^{-1}$. 
Our study contributes to a growing catalog of this rare kind of strong lensing systems and demonstrates the effectiveness of spectroscopic integral field unit observations and advanced modeling techniques in understanding the properties of these systems. 
\end{abstract}



\section{Introduction}

Strong gravitational lensing occurs when a massive object warps space-time and causes the path of light from a well-aligned distant source to bend, typically resulting in multiple images. 
These systems are a powerful tool for astrophysics and cosmology. 
They have been used to study how dark matter is distributed in galaxies and clusters 
\citep[e.g.][]{1991ApJ...373..354K, 2006ApJ...638..703B, 2009ApJ...707L..12H, 
2010Sci...329..924J, 2020Sci...369.1347M} 
and are uniquely suited to probe the low end of the dark matter mass function and test the predictions of dark matter models beyond the local universe \cite[e.g.,][]{2010MNRAS.408.1969V, 2022MNRAS.515.4391S}. 
In addition, multiply lensed quasars and supernovae (SNe) can be used to measure time delays and the Hubble constant, $H_0$ \citep[e.g.,][]{1964MNRAS.128..307R, 2010ARA&A..48...87T, 2020A&A...644A.162S}.

When the alignment is nearly perfect and the lens mass has an elliptical distribution, the background source would appear as quadruply lensed.  These systems are prized as they provide the strongest constraint on the lens mass distribution.
One specific example of quadruply lensed system 
is the Einstein Cross, where four distinct images of the background source form a cross-like pattern with a high degree of symmetry.

The very first Einstein Cross discovered was a lensed quasar system, QSO 2237+0305, also known as Huchra's Lens
\citep{1985AJ.....90..691H}.
This was followed by the discoveries of quite a few other similar systems, some of which 
were used to measure the time-delay $H_0$\citep[e.g.,][]{2020MNRAS.498.1420W}.
Recently, \citet{2021ApJ...921...42S} reported 12 quadruply imaged quasars identified in the Gaia DR2 dataset,
several of them being Einstein Crosses.

In the category of galaxy-galaxy strong lensing, there have been only a handful confirmed cases of Einstein Crosses \citep[e.g.,][]{1995ApJ...453L...5R, 2006ApJ...646L..45B, 2019ApJ...873L..14B, 2020ApJ...904L..31N}. 
In addition to constraining the mass distribution of the lensing galaxies, careful modeling of an increasing number of such systems, covering wider ranges of redshift and mass, also provide valuable mass profile prior for modeling lensed quasars to obtain more accurate $H_0$ measurements \citep[e.g.,][]{2020A&A...643A.165B}.


We report the confirmation of a new galaxy-galaxy lensing Einstein Cross: DESI-253.2534+26.8843 ($\alpha$ = 16:53:00.82, $\delta$ = +26:53:03.48), discovered by \citet{2021ApJ...909...27H} in the Dark Energy Spectroscopic Instrument (DESI) Legacy Imaging Surveys Data Release 8 using deep residual neural networks. 
Out of the 1312 candidates reported in \citet{2021ApJ...909...27H}, this system is one of the 216 most promising (Grade~A) candidates. The brightness of the lens is 23.66 g mag, 21.78 r mag, and 20.35 z mag, and the photometric redshift reported by \citet{2021MNRAS.501.3309Z} is $z_{phot} = 0.649 \pm 0.035$.
Compared with other known galaxy-galaxy lensing Einstein Crosses, this system has the highest redshift (but comparable to the lensed quasar systems used to measure $H_0$) and has the largest Einstein radius.\\

In Section~\ref{sect:MUSEspectra} we present the MUSE observations and the spectroscopic confirmation, in Section~\ref{sect:lenstronomy} we model this system with GIGA-Lens \citep{2022ApJ...935...49G}, and in Section~\ref{sect:summary} we summarize and conclude.

\section{MUSE observations and spectroscopic confirmation}
\label{sect:MUSEspectra}

DESI-253.2534+26.8843 was observed on 2023-05-22 at 04:00h UT as part of a ESO filler program for characterizing of galaxy-galaxy gravitational lensing system candidates (Prog. ID 0111.A-0407, PI Cikota) with the Multi Unit Spectroscopic Explorer (MUSE, \citealt{2010SPIE.7735E..08B}), mounted at UT4 of ESO's Very Large Telescope (VLT) on Cerro Paranal in Chile. 
MUSE is an integral field unit spectrograph with a field of view of 60 arcsec $\times$ 60 arcsec (in the Wide Field Mode) and a spatial resolution of 0.2 arcsec. The spectral range is from 4750 to 9350 \AA\, with the spectral resolution ranging from R = 2000 to 4000 across the
wavelength domain.

The observations were taken with 4 $\times$ 700 seconds exposures during good seeing of $\sim 0.6''$ and thick transparency, and reduced following standard procedures with the MUSE pipeline package version 2.2 \citep{2020A&A...641A..28W} that is a part of the ESO Recipe Execution Tool (ESOREX). We also removed sky lines using the Zurich Atmosphere Purge (ZAP) sky subtraction tool \citep{2016MNRAS.458.3210S}.

\begin{figure*}
\centering
\includegraphics[width=0.4\textwidth]{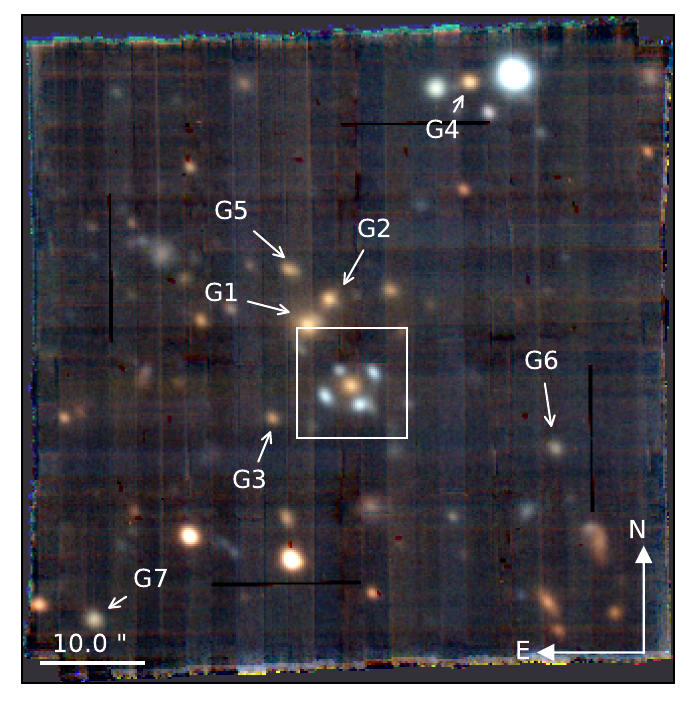}
\includegraphics[width=0.41\textwidth]{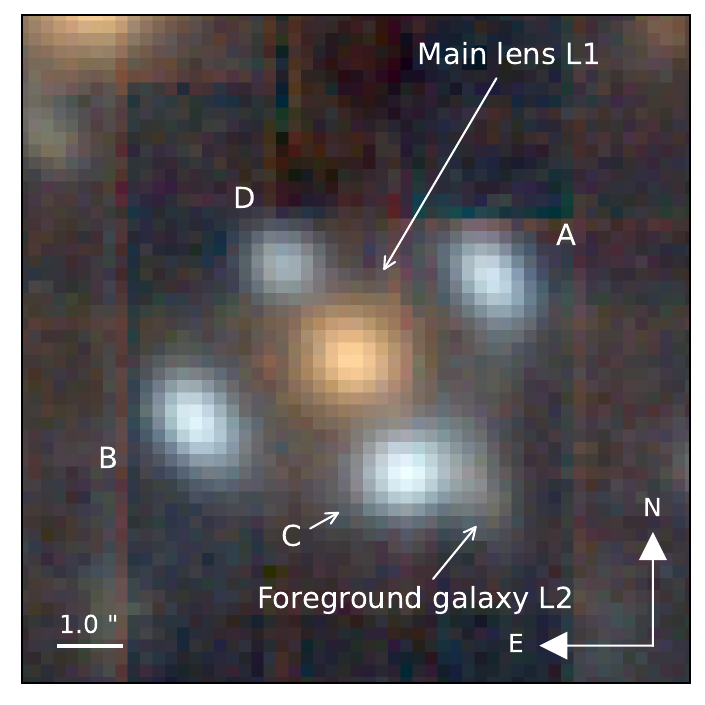}
\caption{\textit{Left panel:} Color image of DESI-253.2534+26.8843 observed with MUSE on May 22nd, 2023. The image is a combination of the SDSS $g$ (blue), SDSS $r$ (green), and SDSS $i$ (red) images generated from the MUSE data cube. Other galaxies with similar redshifts to the lens galaxy are indicated with G1 -- G7. The white square indicates the cutout shown in the right panel. \textit{Right panel:} Cutout with the strong gravitational lens. The images of the source forming the Einstein Cross are indicated with A, B, C and D.}
\label{fig:MUSElensimg}
\end{figure*}

We generated images of the DESI-253.2534+26.8843 in the SDSS $g$, $r$ and $i$ bands by convolving the MUSE data cube with the SDSS $g$, $r$ and $i$ passbands. Figure~\ref{fig:MUSElensimg} shows a color image of the the MUSE field and the gravitational lens, with the source images marked as A, B, C and D. 

We measured the brightness and positions of the lens and source images in the $g$, $r$ and $i$ images with SExtractor \citep{1996A&AS..117..393B}. The zero points for the different bands have been determined based on the brightness of SDSS catalog stars in the field. The positions and magnitudes are listed in Table~\ref{tab:sourcespositions}.

\begin{table}
	\centering
	\tiny
	\caption{\label{tab:sourcespositions} Positions of the gravitational lens and images}
	\begin{tabular}{lccccc} 
	\hline
Object & $\Delta \alpha$ & $\Delta \delta$ & SDSS $g$ & SDSS $r$ & SDSS $i$  \\
 & arcsec$\rm ^a$ & arcsec$\rm ^a$ & mag & mag & mag  \\
\hline
Lens  & 0  & 0    & 23.97 $\pm$ 0.09  & 22.43 $\pm$ 0.05 & 21.12 $\pm$ 0.02 \\
A & 2.21  & 1.22 &  22.91 $\pm$  0.08 & 22.22 $\pm$ 0.05 & 21.97 $\pm$ 0.02 \\
B & -2.49 & -1.02 & 22.65  $\pm$ 0.08 & 21.92 $\pm$ 0.05 & 21.71 $\pm$ 0.02 \\
C & 0.94  & -1.77 &  22.37 $\pm$ 0.08 & 21.60 $\pm$ 0.05 & 21.41 $\pm$ 0.02 \\
D & -1.12 &  1.46 &  23.71 $\pm$ 0.08 & 23.13 $\pm$ 0.05 & 23.02 $\pm$ 0.03 \\
\hline
\end{tabular}\\
{Notes --- $\rm ^a$ Positions are relative to the gravitational lens ($\alpha$ = 16:53:00.82, $\delta$ = +26:53:03.48).}
\end{table}

\begin{figure*}
\centering
\includegraphics[width=\textwidth]{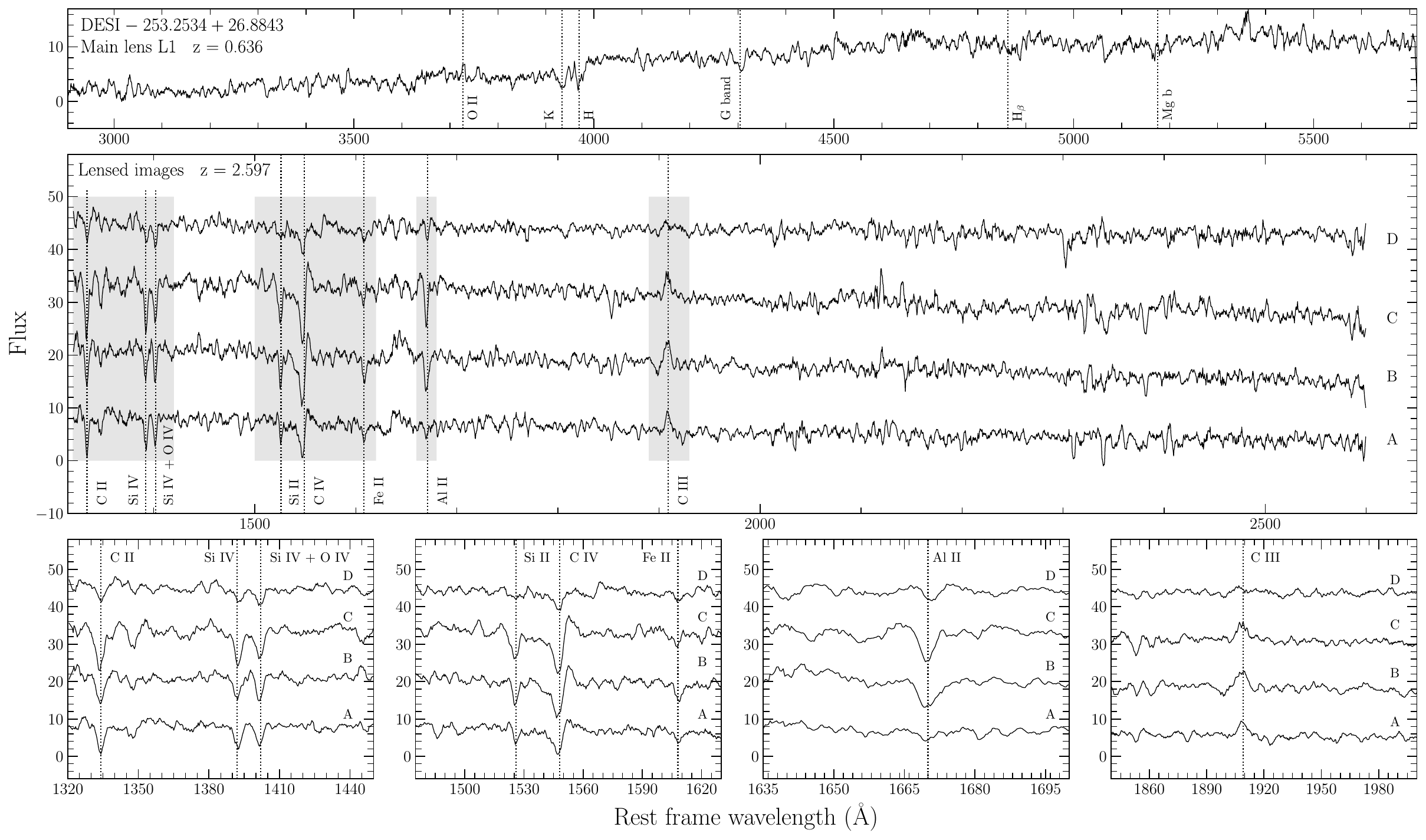}
\caption{\textit{Top panel:} MUSE spectra of the gravitational lens DESI-253.2534+26.8843 in rest frame wavelength at z = 0.636. \textit{Middle panel:} Spectra of the four images of the lensed source in rest-frame wavelength at z = 2.597. \textit{Bottom panels:} Cutouts of the shaded wavelength ranges in the middle panel. We applied the Savitzky-Golay smoothing filter to all spectra using a window size of 9 resolution elements and a first order polynomial.}
\label{fig:MUSEspectra}
\end{figure*}

We extracted the spectra of the lens and the four images of the source from the MUSE data cube and determined the redshifts by manually matching prominent emission and absorption lines\footnote{\href{https://classic.sdss.org/dr6/algorithms/linestable.html}{https://classic.sdss.org/dr6/algorithms/linestable.html}} to the spectra. 
To extract the spectra, we used a circular aperture with a 1'' diameter, which encompassed 13 spaxels, and then calculated average spectra.
The spectra are shown in Figure~\ref{fig:MUSEspectra}. The gravitational lens displays the prominent Fraunhofer absorption lines: Ca II H and K at 3968 \AA\ and 3934 \AA\, respectively, and the G band at 4308 \AA, which undoubtedly places the lens at the redshift of z = 0.636 $\pm$ 0.001.
Furthermore, the spectral energy distribution also shows the 4000 \AA\ break.

All spectra of the four images of the lensed source display consistent spectral features, which confirms that these are indeed lensed images of the same source (Figure~\ref{fig:MUSEspectra}). We identified the $\lambda$ 1336\AA\ C II, $\lambda$ 1394\AA\ Si IV, $\lambda$ 1403\AA\ Si IV, $\lambda$1528\AA\ Si II, $\lambda$1548\AA\ C IV, $\lambda$1609\AA\ Fe II, $\lambda$1670\AA\ Al II absorption lines, and the $\lambda$1909\AA\ C III emission line in the spectra. The spectral characteristics are typical for starbursting galaxies \citep[see e.g. Fig. 4 in][]{1997ApJ...481..673L}.
Based on these spectral features, we are confident that the redshift of the source is at z = 2.597 $\pm$ 0.001.
We note that the image D is fainter compared to the other images, and because of the lower signal to noise ratio (SNR), the spectral features are not as prominent as in the other images.

Based on the color information of the DESI Legacy Surveys data, this  system appears to be embedded in a galaxy group.
We inspected other objects in the MUSE field and found that this is indeed the case. There are 7 additional galaxies with similar redshifts to the lens galaxy (see left panel in Figure~\ref{fig:MUSElensimg}):  
G1 at z = 0.642, G2 at z = 0.641, G3 and G6 at z = 0.636, G4 at z = 0.633, G5 and G7 at z = 0.637. The center of the galaxy group is likely close to G1, G2, G5 and the strong lensing galaxy. 
These galaxies are all passive and display prominent Ca II H and K lines and the 4000 \AA\ break, similar to the spectrum of the main lens L1, except the brightest galaxy G1, which in addition to the Ca II H and K lines and the 4000 \AA\ break also shows a prominent [O II] line and weak H$\beta$ and [O III] lines. A detailed analysis of the galaxy group is out of the scope of this paper. For a discussion on star formation quenching and differences between central and satellite galaxies please refer to e.g. \citet{2015ApJ...800...24K,2018ApJ...860..102W,2019MNRAS.483.5444D} and the references therein.

\begin{figure*}
\centering
\includegraphics[width=1.0\textwidth]{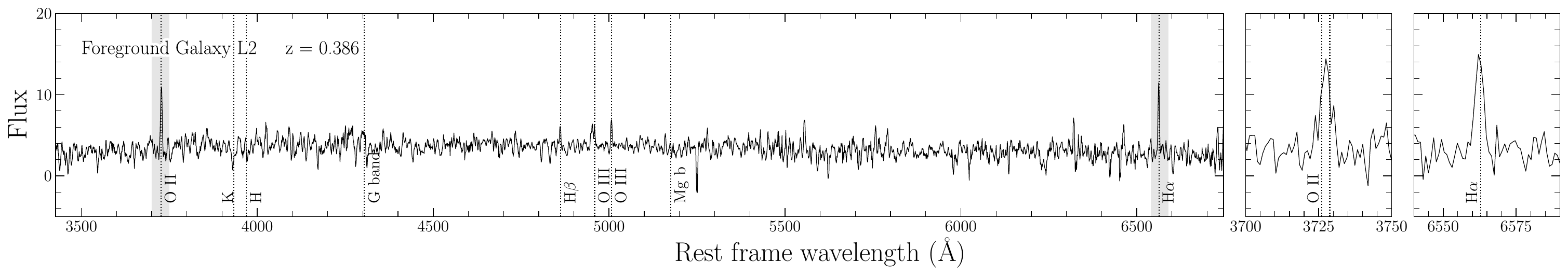}
\caption{Spectrum of the Foreground galaxy \g located in front of the lensed image C at $z = 0.386$. The left panel shows the smooth average spectrum of 15 hand selected spaxels, and the middle and right panels display cutouts of the shaded wavelength ranges around [O II] and H$\alpha$ lines.}
\label{fig:ForegroundGal}
\end{figure*}

Furthermore, we found that in front of the image C, there is a faint foreground galaxy, \g (see Figure~\ref{fig:MUSElensimg}). Figure~\ref{fig:ForegroundGal} shows the spectrum of the galaxy. Although the SNR is lower, there are two prominent emission lines clearly visible, which are consistent with H$\alpha$ and the $\lambda$ $\lambda$3726,3729 \AA\ [O II] doublet at an redshift of $z = 0.386$. While the resolution and signal to noise is not sufficient for resolving the [O II] doublet, other identifiable features in the spectrum correspond to the wavelengths of the H$\beta$ line, [O III] lines, and the Ca II H and K lines.

\section{Analysis}
\label{sect:lenstronomy}

\subsection{Lens modelling with GIGA-Lens}

We use GIGA-Lens \citep{2022ApJ...935...49G} to model this system.  
GIGA-Lens is a Bayesian lens modeling pipeline, 
consisting of three steps: finding the maximum a posteriori (MAP) for the lensing parameters via gradient descent, determining a surrogate multidimensional Gaussian covariance matrix for these parameters using variational inference (VI), and finally sampling with Hamiltonian Monte Carlo.
All three steps use gradient descent with automatic differentiation and take advantage of GPU acceleration.  
It is robust and very fast, typically on the order of minutes to model a system.

Our model comprises a main lens and a second lens (the faint reddish object next to image C, \g).
The masses of both lenses are modeled as singular isothermal ellipsoid (SIE).  
We model the light for both the main lens and the lensed source using the elliptical S\'{e}rsic profile. For the secondary lens light a spherical S\'{e}rsic profile is used.
This model is composed of a total of 31 parameters. All of them are defined in Table~\ref{tab:parameters}. We use the $g$-band image for modeling and a 15$\times$15 pixel Gaussian with FWHM of $0.6''$ as our PSF.

\begin{table}
\caption{Prior distribution used for lens modeling}
\begin{align*}
\text{Lens Mass:} &
\begin{cases}
    \hfill \theta_E & \sim \exp\left(\mathcal{N}(\textcolor{blue}{\ln 2, 0.25} / \textcolor{red}{\ln0.3, 0.25})\right) \\[3pt]
    \hfill \epsilon_{1}, \epsilon_{2} & \sim \mathcal{N}(\textcolor{blue}{0, 0.1} / \textcolor{red}{0, 0.1}) \\[3pt]
    \hfill x & \sim \mathcal{N}(\textcolor{blue}{0, 0.05} / \textcolor{red}{2, 0.1}) \\[3pt]
    \hfill y & \sim \mathcal{N}(\textcolor{blue}{0, 0.05} / \textcolor{red}{-2.4, 0.1}) \\[3pt]
    \gamma_{ext, 1}, \gamma_{ext, 2} & \sim \mathcal{N}(0, 0.05)
\end{cases} \\
\text{Lens Light:} &
\begin{cases}
    \hfill R_l & \sim \exp\left(\mathcal{N}(\textcolor{blue}{\ln 1, 0.15} / \textcolor{red}{\ln0.5, 0.15})\right) \\[3pt]
    \hfill n_l & \sim \mathcal{U}(\textcolor{blue}{1, 5} / \textcolor{red}{1, 5}) \\[3pt]
    \hfill \epsilon_{l,1}, \epsilon_{l,2} & \sim \mathcal{TN}(\textcolor{blue}{0, 0.1, -0.3, 0.3} / \textcolor{red}{-}) \\[3pt]
    \hfill x_l & \sim \mathcal{N}(\textcolor{blue}{0, 0.05} / \textcolor{red}{2.0, 0.1}) \\[3pt]
    \hfill y_l & \sim \mathcal{N}(\textcolor{blue}{0, 0.05} / \textcolor{red}{-2.4, 0.1}) \\[3pt]
    \hfill I_l & \sim \exp\left(\mathcal{N}(\textcolor{blue}{\ln25.0, 0.3} / \textcolor{red}{\ln25.0, 0.3})\right)
\end{cases} \\
\text{Source Light:} &
\begin{cases}
    \hfill R_{\text{s}} & \sim \exp\left(\mathcal{N}(\ln 0.25, 0.15\right) \\[3pt]
    \hfill n_s & \sim \mathcal{U}(0.5, 4) \\[3pt]
    \hfill \epsilon_{s,1}, \epsilon_{\text{s},2} & \sim \mathcal{TN}(0, 0.15, -0.5, 0.5) \\[3pt]
    \hfill x_s & \sim \mathcal{N}(0, 0.25) \\[3pt]
    \hfill y_s & \sim \mathcal{N}(0, 0.25) \\[3pt]
    \hfill I_s& \sim \exp\left(\mathcal{N}(\ln 150.0, 0.5)\right)
\end{cases}
\end{align*}
 
\label{tab:parameters}

{\small Notes --- The model consists of SIE for both lens mass profiles and external shear. 
$\theta_E$ is the Einstein radius in arcsec.  $x$ and $y$ are the mass centers of the two lenses. 
$\gamma_{1, \text{ext}}$ and $\gamma_{1, \text{ext}}$ are the external shear.
The parameters $\epsilon_1$ and $\epsilon_2$ are the lens mass eccentricities, while $\epsilon_{l,1}$, $\epsilon_{l,2}$ and $\epsilon_{s,1}$, $\epsilon_{s,2}$ are the lens and source light eccentricities, respectively.
We employ the notation $\textcolor{blue}{a} / \textcolor{red}{b}$ to indicate the parameters for the \textcolor{blue}{main lens}/\textcolor{red}{secondary lens}. 
We use a spherical S{\'e}rsic profile for the light of the secondary lens, for which eccentricity priors are thus indicated with a dash.
$R_l$, $R_s$ are defined as the half-light radius and $n_l$, $n_s$ as the Sérsic index. $x_l$, $y_l$ and $x_s$, $y_s$ describe the center of the light and $I_l$, $I_s$ its intensity. Subscripts $l$, $s$ imply the parameter belongs to the lens light profile or to the source light profile, respectively.
Here, $\mathcal{U}(a, b)$ is a uniform distribution with support $[a, b]$, $\mathcal{N}(\mu, \sigma)$ is Gaussian with mean $\mu$ and standard deviation $\sigma$, and $\mathcal{TN}(\mu, \sigma; a, b)$ is a truncated Gaussian with support $[a, b]$. }
\end{table}

\begin{figure*}
    \centering
    \includegraphics[scale = 0.52]{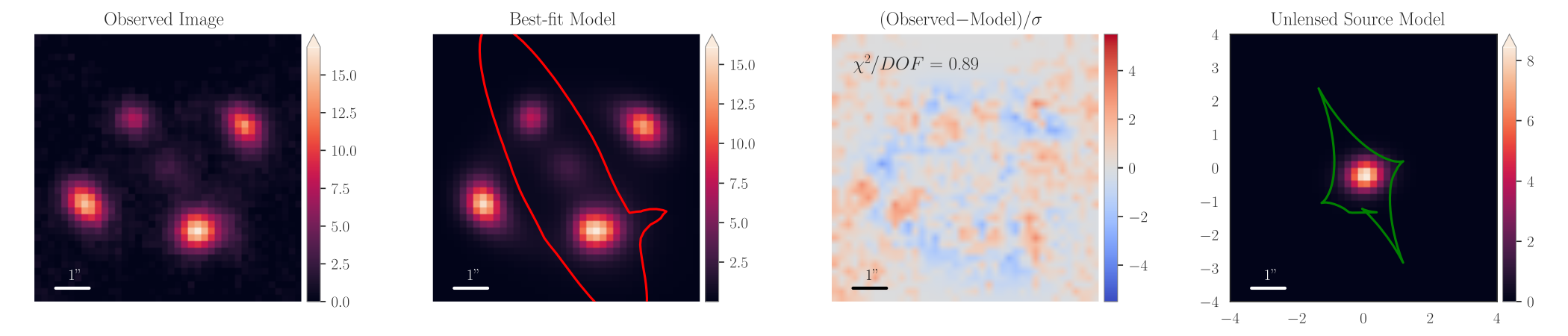}
    \caption{The results for our model consisting of a main lens and a second lens (\g, at $z = 0.386$). From left to right, we show: the observed MUSE image in $g$-band (North is up, East is left), the best-fit model with the critical curve, the reduced residual, and the unlensed source with the caustic.}
    \label{fig:model}
\end{figure*}

We achieved very good residual and excellent sampling results. 
We report two metrics that are widely used in the statistics literature to measure the degree of convergence of our sampler: the potential scale reduction factor (PSRF), $\hat{R}$ \citep{1992StaSc...7..457G} and the effective sample size (ESS).
The $\hat{R}$ values for nearly all parameters are below 1.1, with the lens light eccentricity the only exception (for $\epsilon_{\text{l}2}$, $\hat{R} = 1.2$).
Given how faint it is, that is hard to constrain. For the same reason, the only two parameters whose effective sample sizes (ESS) are below 2000 are $\epsilon_2$ and $n_{sersic}$ of the lens light. 
The maximum ESS is around 24000.
Our best-fit model 
is shown in Figure~\ref{fig:model}, 
with mass parameters presented in Table \ref{tab:model_comp}.

\begin{table}
    \caption{Best-fit mass parameters for the main lens \G, the secondary lens \g, and the external shear. }
    \renewcommand{\arraystretch}{0.4}
    \begin{tabular}{ccc}
         Parameters & Main Lens, \G & Secondary Lens, \g \\ \hline\hline
         \\ $\theta_E$ & $2.520_{-0.031}^{+0.032}$  & $0.261_{-0.027}^{+0.028}$  \\ \\
         $\epsilon_1$ & $-0.365_{-0.009}^{+0.009}$ & $0.662_{-0.069}^{+0.069}$  \\ \\
         $\epsilon_2$ & $-0.486_{-0.011}^{+0.011}$ &  $0.304_{-0.062}^{+0.064}$ \\ \\
         $x$          & $-0.115_{-0.011}^{+0.010}$  & $1.836_{-0.107}^{+0.096}$ \\ \\
         $y$          & $-0.207_{-0.018}^{+0.018}$   & $-1.563_{-0.059}^{+0.073}$ \\ \\ \hline
         \\ $\gamma_{1, \text{ext}}$ & \multicolumn{2}{c}{$-0.008_{-0.005}^{+0.006}$ } \\ \\
         $\gamma_{2, \text{ext}}$ & \multicolumn{2}{c}{$-0.038_{-0.006}^{+0.006}$ }  \\ \\ \hline 
    \end{tabular}
    \label{tab:model_comp}
    
\end{table}

\begin{figure*}
  \centering
  \vspace{0.2cm}
  \includegraphics[scale = 0.42]{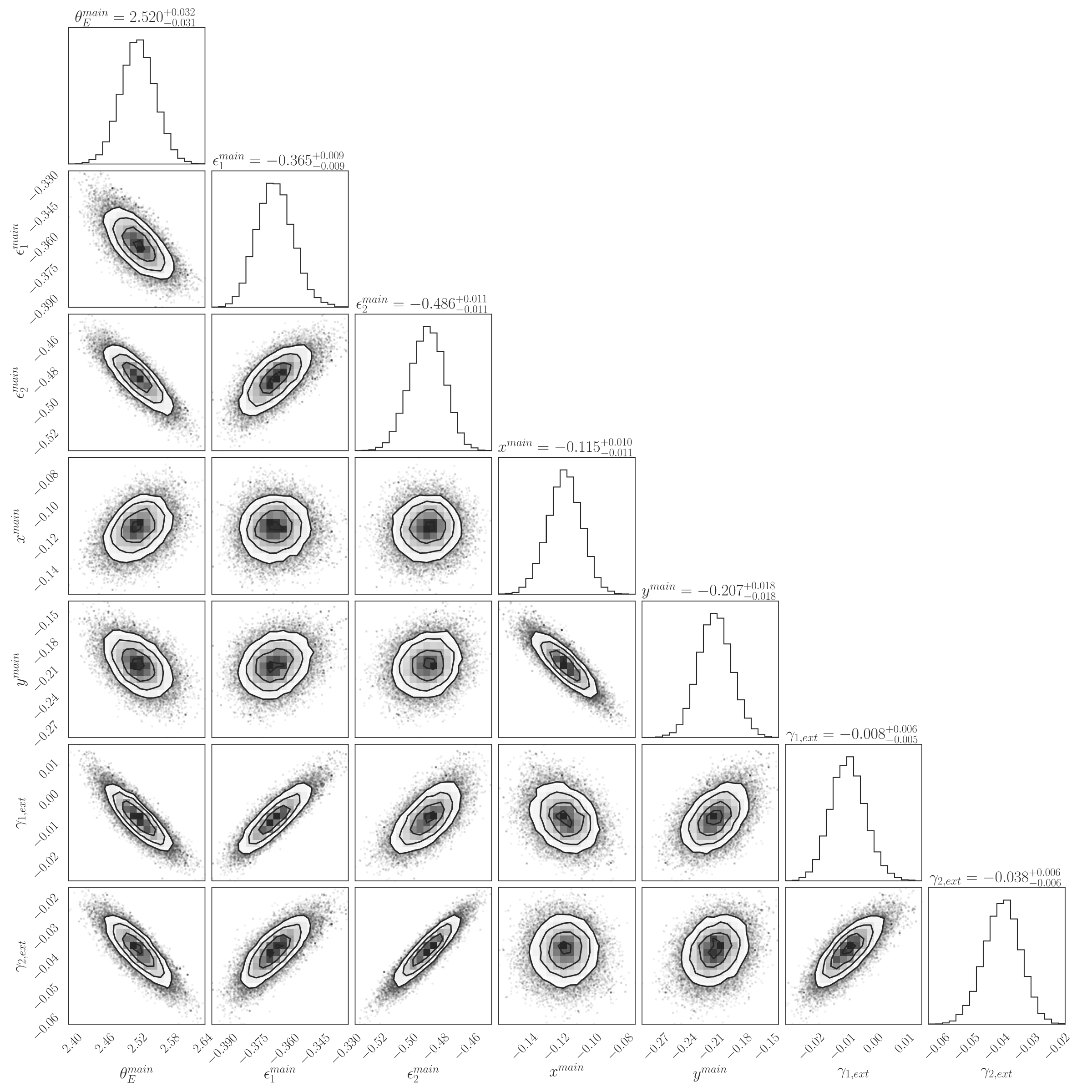}
  \begin{tikzpicture}[remember picture,overlay]
      \node at (-5.5cm,15.0cm) 
      {\includegraphics[scale = 0.47]{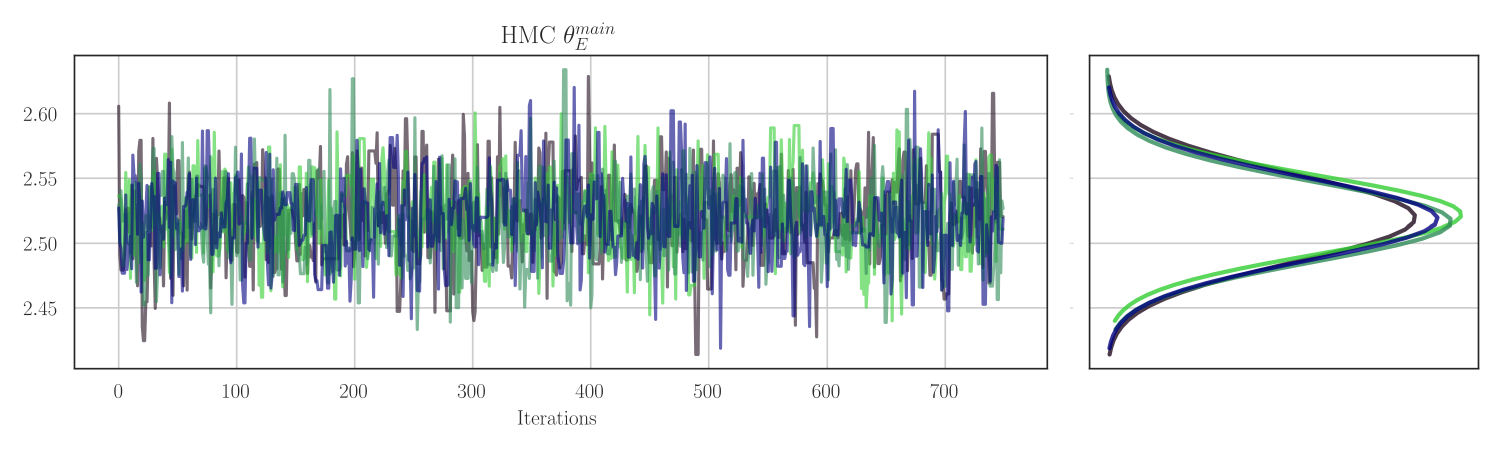}};
    \end{tikzpicture}
  \caption{The corner plot shows our sampling results for the main lens and external shear mass parameters. 
  In the inset, four chains of HMC demonstrate statistical consistency within the sampling of the Einstein radius, with $\hat{R} = 1.007$ (see text).}
  \label{fig:cornerplot}
\end{figure*}

We show representative sampling results for the main lens 
along with individual chains for the Einstein radius in Figure~\ref{fig:cornerplot}. The sampling for all other parameters are equally good.

For the main lens, we determined the Einstein radius to be $\theta_{E} ={2.520''}_{-0.031}^{+0.032}$ (Table~\ref{tab:model_comp}). 
This can be converted to the velocity dispersion as follows \citep[e.g.,][]{1996astro.ph..6001N}:

\begin{equation}\label{eqn:vdisp}
    \theta_{E} = 4 \pi \left( \frac{\sigma_{\rm SIE}}{c} \right) ^2 \frac{D_{\G\textnormal{-}s}}{D_s} ,
\end{equation}
where $D_{\G\textnormal{-}s}$ and $D_s$ are the angular-diameter distances between the main lens and the source and from the observer to source, respectively.
Using the cosmological parameters from
\citet{2020A&A...641A...6P}, with $H_0$ = 67.4 $\pm$ 0.5 km s$^{-1}$ Mpc$^{-1}$, $\Omega _{\Lambda}$ = 0.6847 $\pm$ 0.0073, and $\Omega _{M}$ = 0.315 $\pm$ 0.007, 
we obtained $D_s$ = 1690.6 Mpc and $D_{\G-s}$ = 1026.1 Mpc, respectively.
This led to a velocity dispersion of $\sigma^{\mathrm{L1}}_{\rm SIE} = 379 \pm 2\, \rm{km \, s}^{-1}$ for the main lens. 
This is among the most massive galactic scale strong lenses \citep[e.g.,][]{2015ApJ...811...20C, 2019ApJ...873L..14B}.

The predicted magnifications are 2.40, 3.23, 4.03 and 0.81 for images A, B, C and D, respectively, with the total being 10.47.

We now briefly discuss the secondary lens, which is the foreground galaxy \g at $z = 0.386$, located near (in projection) the image C.
Our modeling approach is informed by the following: (i) the data we use for lens modeling is from low-resolution ground-based observations; (ii) the main lens is much more massive than the secondary lens; and (iii) the secondary lens is near the Einstein ring of the main lens.
Instead of pursuing lens modeling with double lens planes, we 
treat this interloper as an ``effective'' subhalo of the main lens 
\citep{PhysRevD.102.063502}.
 We determine its effective Einstein radius to be $0.261{''}^{+0.028}_{-0.027}$ (Table~\ref{tab:model_comp}).
By treating \g as an ``effective'' subhalo,
its net lensing effect of the foreground galaxy \g is reduced by a factor of $1 - \beta$ \citep{PhysRevD.102.063502}, with $\beta = \frac{D_{\g\textnormal{-}\G} D_s}{D_{\G} D_{\g\textnormal{-}s}}$, 
where $D_{\g \textnormal{-}\G}$ and $D_{\g\textnormal{-}s}$ are the angular diameter distances between the two lenses and between the secondary lens and the source, respectively (\citealt{PhysRevD.102.063502}, see also \citealt{2021JCAP...08..024F}).
In our case, $\beta = 0.470$.  
Thus, the lensing effect has been reduced by 0.53.  
Given the SIE mass model, we adjusted the value implied by Eq~(\ref{eqn:vdisp}) accordingly and determined  
$\sigma_\mathrm{SIE}^\mathrm{L2}$ 
to be $110\, \mathrm{km \, s}^{-1} / \sqrt{0.53} = 137\, \mathrm{km \, s}^{-1}$.
This indicates that it is a relatively low-mass galaxy (see, e.g., \citealt{2015ApJ...811...20C}), and  
 is consistent with its low brightness.

In recent years there has been a growing interest in using gravitational imaging \citep{2005MNRAS.363.1136K} to detect sub-galactic scale halos with no baryons ($M_{halo} \lesssim 10^9 M_{\odot}$). 
Sometimes referred to as ``dark halos'', they include both subhalos and line-of-sight (LOS) interlopers.  
This potentially provides a powerful way to test dark matter models \citep[e.g.,][]{2010MNRAS.408.1969V, 2018MNRAS.475.5424D}.  
In this approach, the detection of a dark halo and the measurement of its mass solely rely on lens modeling. 
DESI-253.2534+26.884 is a rare system with (i) a main lens that can be well constrained (due to the Einstein Cross pattern) and (ii) a small \emph{visible} LOS interloper near (in projection) the Einstein ring of the main lens.
Such a system can provide an important check for the gravitational imaging approach for detecting and measuring the mass of LOS \emph{dark} halo interlopers\footnote{For the detection of subhalos using the gravitational imaging technique, \citet{2010MNRAS.407..225V} performed testing by applying it to SDSSJ120602.09+514229.5, which has a visible satellite galaxy (hosted by a subhalo) at the location of the Einstein ring formed by the main lens. But to our knowledge, such tests have not been done for the case of interlopers.
}.
By obtaining high resolution imaging for this system, as well as resolved dynamical information for the main lens and the interloper (e.g., VLT/MUSE with adaptive optics),  
we can check the agreement between the mass measurements from careful lens modeling and detailed dynamics.
This can be used to establish the validity of using gravitational imaging to detect and measure the mass of LOS sub-galactic dark halos, which some groups claimed are better than subhalos to determine the nature of dark matter \citep[e.g.,][]{2022MNRAS.515.4391S}.

\subsection{pPXF fitting}
\label{sect:pPXT}

We fit the lens galaxy spectrum with the pPXF package (\citealt{Cappellari2022}, see also \citealt{2004PASP..116..138C} and \citealt{2017MNRAS.466..798C}) with the goal to determine the stellar velocity dispersion. Following the pPXF usage examples\footnote{\href{https://github.com/micappe/ppxf_examples}{https://github.com/micappe/ppxf\_examples}} for high redshift galaxies, the spectrum was fitted using E-MILES stellar population synthesis models \citep{2016MNRAS.463.3409V}. 
The most prominent features of our spectrum are the Ca II H and K lines, and we obtain a best-fit velocity dispersion of $\sigma$ = 403 $\pm$ 85 km s$^{-1}$.
This is in good agreement with the velocity dispersion from lens modeling.

\section{Summary and conclusion}
\label{sect:summary}

We present MUSE observations of the strong gravitational lens system DESI-253.2534+26.884, which was discovered in the DESI Legacy Imaging Surveys data using deep residual neural networks by \citet{2021ApJ...909...27H} and appeared as an Einstein cross. 
We determined the redshift of the main lensing galaxy, $z_{L1} = 0.636 \pm$ 0.001, and show that the four images of the source display common spectroscopic features which places the source at a redshift of $z_s = 2.597 \pm 0.001$, fully confirming this to be a strong lensing system.
We found a faint foreground galaxy located in front of the image C (see Figure~\ref{fig:MUSElensimg}). A careful selection of the spaxels allowed us to extract the spectrum and determine its redshift to be $z_{L2} = 0.386$. This set of redshifts especially demonstrate the advantage IFU observations of gravitational lens systems in contrast to long slit spectroscopy.

We modeled the gravitational lens system using GIGA-Lens \citep{2022ApJ...935...49G}. The Einstein radius of the lens is $\theta_{E} =2.520{''}_{-0.031}^{+0.032}$, which corresponds to a velocity dispersion of $\sigma^{L1}_{\rm SIE}$ = 379 $\pm$ 2 km s$^{-1}$. This is consistent with the spectroscopically determined velocity dispersion of $\sigma$ = 403 $\pm$ 85 km s$^{-1}$.

To our knowledge, this is the first time a real gravitational lensing system has been modeled with GPUs, using the GIGA-Lens pipeline.
And it shows great promise:  The modeling time is $55$~sec on a single A100 GPU on the Perlmutter supercomputer at the National Energy Research Scientific Computing Center (NERSC). 
With 4 GPUs on one GPU node, we expect the time to be roughly $15$~sec \citep{2022ApJ...935...49G}.  
By comparison, for ground-based data from DES, \citet[][R21]{2021arXiv210900014R} reported an average lens modeling time of 4.3 hour using Lenstronomy (e.g., \citealt{2018PDU....22..189B}).  
Our model is comparable in that the main lens is modeled as SIE just as in R21.  
In our model, due to the presence of the second lens, we have more parameters than R21. 
Yet, we have achieved greater than two orders of magnitude speedup.
This concretely demonstrates a very promising future of modeling $\mathcal{O}(10^5)$ of strong lensing systems \citep[e.g.,][]{2015ApJ...811...20C} that are expected to be discovered in the next decade (e.g., Euclid, LSST, and the Roman Space Telescope), in a fast, robust and scalable way.


\section*{Acknowledgments}
\small
We thank Greg Aldering, Adam Bolton, Saul Perlmutter, and Yiping Shu for insightful discussion.
The work of A.C. is supported by NOIRLab, which is managed by the Association of Universities for Research in Astronomy (AURA) under a cooperative agreement with the National Science Foundation. 
X.H. acknowledges the University of San Francisco Faculty Development Fund.
This work was supported in part by the Director, Office of Science, Office of High Energy Physics of the US Department of Energy under contract No. DE-AC025CH11231.
This research used resources of the National Energy Research Scientific Computing Center (NERSC), a U.S. Department of Energy Office of Science User Facility operated under the same contract as above and the Computational HEP program in The Department of Energy's Science Office of High Energy Physics provided resources through the ``Cosmology Data Repository'' project (grant No. KA2401022).
This work is based on observations collected at the European Organisation for Astronomical Research in the Southern Hemisphere under ESO program 0111.A-0407. The execution in the service mode of these observations by the VLT operations staff is gratefully acknowledged.

\bibliography{impolbib}{} 
\bibliographystyle{aasjournal}



\end{document}